\documentclass[11pt,a4paper]{article}
\pdfoutput=1
\usepackage{jheppub}
\usepackage{graphicx} 
\usepackage{amsmath,amssymb}
\usepackage{amsfonts}
\usepackage{url}
\usepackage{float}
\usepackage[toc,page]{appendix}
\usepackage{lipsum}
\title{Entanglement of Vacuum States With the de Sitter Horizon: Consequences on Holographic Dark Energy}
\author{Rafael Pav\~ao,}
\author{Ricardo Faleiro,}
\author{Alex H. Blin}
\author{and Brigitte Hiller}
\affiliation{CFisUC, Department of Physics, University of Coimbra, 3004-516 Coimbra, Portugal}
\emailAdd{rafaelpppavao@gmail.com}
\emailAdd{ricardoandremiguel@hotmail.com}
\emailAdd{alex@uc.pt}
\emailAdd{brigitte@teor.fis.uc.pt}
\abstract{
	The aim of this article is to study the effect of an Event Horizon on the entanglement of the Quantum Vacuum and how entanglement, together with the Holographic Principle, may explain the current value of the Cosmological Constant, in light of recent theories.  Entanglement is tested for vacuum states very near and very far from the Horizon of a de Sitter Universe, using the Peres-Horodecki (PPT) criterion. A scalar vacuum field ($\hat{\phi}$) is averaged inside two boxes of volume $V$ in different spatial positions such that it acquires the structure of a bipartite Quantum Harmonic Oscillator, for which the PPT criterion is a necessary but not sufficient condition of separability.
	Entanglement is found between states obtained from boxes shaped as spherical shells with thickness of the order of one Planck distance ($l_p$), when one of the states is near the Horizon, and the other state is anywhere in the Universe. Entanglement disappears when the distance of the state near the horizon and the Horizon increases to around $5l_p$. If we consider the Horizon not as a surface but as a spherical shell of thickness $l_p$, then this means that there is entanglement between the states in the Horizon and the rest of the Universe. 
	When both states are at distances larger than $\sim 5 l_p$ from the Horizon, no entanglement is found.
}
\arxivnumber{1607.02115}

\begin{document}
\maketitle
\flushbottom

\section{Introduction}

In 1998, evidence was presented suggesting that our Universe is expanding at an accelerating rate, indicating the presence of Dark Energy \cite{ccev,ccev2}. Since then, numerous attempts have been made to explain the physical meaning of such energy, but up to date, no complete explanation has been found for the phenomenon.

One of the first and well-known tentatives to explain the Cosmological Constant is based on the energy of the vacuum of Quantum Field Theory up to the Planck scale. The resulting value of the Cosmological Constant turns out, however, to be 123 orders of magnitude greater than the observed value, which is the notorious Cosmological Constant problem.

Several alternatives based on the vacuum energy can be found in the literature\cite{oandn,oandn12,oandn22}. None of these explanations is, however, fully established, and in many cases they are not in agreement with the observed data\cite{oandn}.

A few years before the evidence for the Cosmological Constant was found, G. 't Hooft \cite{thooft} and L. Susskind \cite{susskind1} noted that the number of degrees of freedom in our universe is more restricted than usually assumed. The universe must have entropy that does not surpass the entropy of a Black Hole of the same size. Such entropy grows with the area of its boundary, not with its volume. This means that the information inside a physical system is not proportional to its volume, but, at the most, to its surface. Our 3-dimensional Universe must have the same amount of degrees of freedom as the 2-dimensional surface that surrounds it. This is called the Holographic Principle.

Therefore one faces the problem that the vacuum energy of our Universe should not be calculated  just by counting every state inside its volume, since one would be over counting the number of states allowed by the Holographic Principle. The question is then, if there is some dilution of states as consequence of the Holographic Principle, how exactly should one obtain the right amount of states? Some attempts were made with a good degree of success \cite{saulo,kimlee,capoz1,capoz2,capoz3}, however they usually rely on using a vacuum energy that is chosen to fit the holographic bound, or choosing a certain energy cutoff that is not necessarily natural.

It would be thus most desirable to have a procedure that allows to calculate the vacuum energy in a way that respects the Holographic Principle, without the need to introduce an unnatural energy cutoff.

Several papers \cite{weinberg1,marks,shinji,bombelli,muller,wolf} suggest that the entropy of entanglement of a scalar field in the presence of a spherical boundary (be it a horizon, a physical boundary, or an imaginary one) is proportional to the area of that boundary, just like in the Holographic Principle. In \cite{marks,shinji,bombelli,muller} entanglement was detected by constructing the density matrix of a scalar field $\hat{\phi}$ in a lattice with a certain spacing\footnote{This approximation is used because a spatial cutoff (such as the Planck length) produces a sort of lattice.} and then tracing out the states outside (or inside) the boundary in order to calculate the von Neumann entropy. In this way, the field could be discretized, facilitating the tests for the presence of entanglement. In \cite{weinberg1} the possibility is suggested that entanglement might explain the entropy of a Black Hole, and in \cite{kimlee} it is suggested that it is the energy of the entanglement itself, originated by the spherical boundary of our visible Universe, that creates the accelerated expansion of the Universe. This latter theory gives the same result as the holographic Cosmological Constant theories. Originally stated in the context of Black Hole physics, the Holographic Principle  is found in lattice simulations to be also at work in a variety of classical and quantum systems. In thermal equilibrium these exhibit  an area law when the correlation length, measured in terms of the mutual information, is finite \cite{wolf}.

These works point at the physical significance that correlations play in the determination of area laws. In particular, the role played by quantum entanglement in the Holographic Principle must be assessed in order to attribute a holographic nature to dark energy.

To try to shed light on this question, we propose the study of entanglement between vacuum states in a de Sitter Universe, a Universe dominated by the vacuum energy, to which our present Universe will evolve in the future. To be able to calculate entanglement of a scalar field $\hat{\phi}$, which is a continuous variable system, we shall consider the field averaged inside two boxes (each centered at different points in space) such that the problem is reduced  to the study of bipartite Quantum Harmonic Oscillators (each box is a state)\cite{vogel}. In this way, we will be able to use the PPT criterion to test for the existence of entanglement between vacuum states in the de Sitter Universe, especially  between vacuum states and the de Sitter Horizon. The Horizon states will be taken as the states inside a spherical shell of Planck length thickness \cite{weinberg1,pavao}.

Finally, it should be noted that, in this case, results may be different from the ones in \cite{weinberg1,marks,shinji,bombelli,muller,wolf}, since the Negativity and the Entropy of entanglement are both measures of entanglement, but are not equivalent. Also, we consider entanglement between two states inside a boundary (in this case, the de Sitter Horizon) and not entanglement between regions of space separated by a boundary. Moreover, contrary to \cite{marks,shinji,bombelli,muller}, we use a continuous field $\hat{\phi}$.

The structure of this article is as follows.
In section 2 we present the method used to detect the entanglement in a de Sitter universe, in section 3 we show the numerical results obtained from the method and provide direct interpretations, and in section 4 we discuss a more general interpretation of the results along with some possible improvements and continuations to our work.

\section{Entanglement of States With the Horizon}

To test for the entanglement between $\hat{\phi}(x)$ and $\hat{\phi}(x')$ a method was developed in \cite{kofler} that allows the use of the PPT criterion in quantum field theory. The method consists in averaging the field inside a box centered at $x$ and ignoring everything outside the box, simulating in this way a hypothetical measurement of the field in $x$ with a certain spatial definition. Thus we can define two field modes,
\begin{equation}
\label{averagestates}
\left\{\begin{matrix}
\hat{\Phi}_B(\mathbf{r},t) = \frac{1}{V}\int_B d \mathbf{y} \ \hat{\phi}(\mathbf{r}+\mathbf{y},t), \ \ \ \ \ \ \\
\\
\hat{\Phi}_{B'}(\mathbf{r}',t') = \frac{1}{V'}\int_{B'}  d \mathbf{y} \ \hat{\phi}(\mathbf{r}'+\mathbf{y},t'),
\end{matrix}\right.
\end{equation}
where $B$ (or $B(\mathbf{r},\mathbf{y})$ if it depends on the position) represents the box being used and $V$ is the volume of said box.
These modes form a bipartite system of quantum harmonic oscillators.

In \cite{rsimon} it was proved that, for a bipartite system of gaussian states defined by a vector,
\begin{equation}
\boldsymbol{\hat{\xi}} = (\hat{x}_1,\hat{p}_1,\hat{x}_2,\hat{p}_2)^T,
\end{equation}
and a covariance matrix,
\begin{equation}
\label{eq:CM}
\mathbf{V}_{\alpha \beta} = \left[ \begin{pmatrix}
\mathbf{A} & \mathbf{C} \\ 
{\mathbf{C}}^T & \mathbf{B}
\end{pmatrix} \right]_{\alpha \beta} = \frac{1}{2} \left< \left \{ \boldsymbol{\hat{\xi}}_{\alpha}, \boldsymbol{\hat{\xi}}_{\beta} \right \} \right>,
\end{equation}
(where $\hat{x}_i$ and $\hat{p}_i$ are the canonical variables of the respective subsystems and the brackets $<>$ denote the vacuum expectation value) the PPT criterion for separability reduces to
\begin{equation}
\label{simoncrit}
F = \textup{Det}(\mathbf{A})+\textup{Det}(\mathbf{B}) - 2 \textup{Det}(\mathbf{C}) - \frac{1}{4}-4\textup{Det}(\mathbf{V}) \leq 0.
\end{equation}
The idea in \cite{kofler} is that, if $\hat{\Phi}_B(\mathbf{r},t)$ and $\hat{\Phi}_{B'}(\mathbf{r}',t')$ are gaussian states, then a covariance matrix can be constructed using
\begin{equation}
\boldsymbol{\hat{\xi}} = (\hat{\Phi}_B,\hat{\Pi}_B,\hat{\Phi}_{B'},\hat{\Pi}_{B'})^T,
\end{equation}
with $\hat{\Pi}_B = \partial_t \hat{\Phi}_B$, and with the use of \eqref{simoncrit} test for the presence of entanglement. The states are gaussian if the following uncertainty condition is fulfilled \cite{gsbeyond,alessiothesis}:
\begin{equation}
\label{uncrt}
\mathbf{V} + \frac{i}{2} \Omega \geq 0,
\end{equation}  
where
\begin{equation}
\Omega = \begin{pmatrix}
0 & 1 & 0 & 0 \\ 
- 1 & 0 & 0 & 0 \\ 
0 & 0 & 0 & 1 \\ 
0 & 0 & -1 & 0
\end{pmatrix}.
\end{equation}
However, even if the states in \eqref{averagestates} are not gaussian states, \eqref{simoncrit} can still be used as a necessary but not sufficient condition of separability \cite{vogel}.\footnote{To our best knowledge, at present no method to determine uniquely the separability of continuous variable non-gaussian states exists.}
In \cite{libby} this method was extended, proposing that other averaging procedures could be used in \eqref{averagestates} to simulate other types of possible detectors, such that we end up with field modes of the sort
\begin{equation}
\hat{\Phi}_B (x) = \int d y \ g_B (x,y) \hat{\phi}(x,y),
\end{equation}
where $g_B (x,y)$ is called the detector profile.
For simplicity we will use the following $g_B (x,y)$ in \eqref{averagestates}:
\begin{equation}
g_B (x,y) = \frac{1}{V} \times \begin{cases}
1 & \text{ if } \mathbf{r} + \mathbf{y} \in B\\ 
0 & \text{ if } \mathbf{r} + \mathbf{y} \notin B
\end{cases}.
\end{equation}

Since the integrals in \eqref{averagestates} can become difficult to solve, in \cite{kimbound} it was shown that, at least for cubic boxes with $V = L^3$ (where $L$ is the size of the box),
\begin{equation}
\label{inffield1}
\lim_{L \rightarrow 0} \hat{\Phi}_{B}(x) = \hat{\phi}(x),
\end{equation}
and,
\begin{equation}
\label{inffield2}
\lim_{L \rightarrow 0} \hat{\Pi}_{B}(x) = V \hat{\pi}(x),
\end{equation}
where $\hat{\pi}(x) = \partial_t \hat{\phi}(x)$. Equations \eqref{inffield1} and \eqref{inffield2} are valid as long as $L > 0$ (that is, an infinitesimal but not zero volume $V$).

Then, since we will be calculating entanglement between vacuum states, the calculations can be simplified by relating the Hadamard function,
\begin{equation}
\label{hadamard}
H(x,x') = \left< 0 \right|\left \{ \hat{\phi}(x), \hat{\phi}(x') \right \} \left| 0 \right>,
\end{equation}
with the covariance matrix of the system \cite{kimbound}, such that, using the definition in \eqref{eq:CM},
\begin{subequations}
\begin{align}
& \mathbf{A}=\frac{1}{2}\begin{pmatrix}
\lim_{x' \rightarrow x} H(x,x') & V \lim_{x' \rightarrow x} \partial_{t'}H(x,x')\\
V \lim_{x' \rightarrow x} \partial_{t'} H(x,x') & V^2 \lim_{x' \rightarrow x} \partial_t \partial_{t'}H(x,x')
\end{pmatrix},\\
& \mathbf{B} =\frac{1}{2}\begin{pmatrix}
\lim_{x \rightarrow x'} H(x,x') & V' \lim_{x \rightarrow x'} \partial_{t'}H(x,x')\\ 
V' \lim_{x \rightarrow x'} \partial_{t'} H(x,x')  &  {V'}^2 \lim_{x \rightarrow x'}\partial_t \partial_{t'} H(x,x') 
\end{pmatrix},\\
& \mathbf{C}=\frac{1}{2}\begin{pmatrix}
 H(x,x') & V' \partial_{t'} H(x,x')\\
 V \partial_t H(x,x') & VV' \partial_t \partial_{t'} H(x,x')
\end{pmatrix}.
\end{align}
\end{subequations}

Now we need to calculate the field $\phi(x)$ in de Sitter spacetime. From \cite{birrell,tedcurvedfields} we have that, in a curved spacetime,
\begin{equation}
\label{scalfield}
\hat{\phi}(x) = \int d \mathbf{k} \ \left ( u(x,k) \hat{a}_{\mathbf{k}}+ u^*(x,k) \hat{a}^{\dagger}_{\mathbf{k}} \right ),
\end{equation}
where $u(x,k)$ are the generalized field modes for curved spacetimes. We can eliminate the creation/annihilation operators by writing \eqref{hadamard} in terms of the field modes such that
\begin{equation}
H(x,x') = \int d \mathbf{k} \int d \mathbf{k}' \left(u(x,k)u^*(x',k') +u^*(x,k)u(x',k') \right).
\end{equation}
The modes for this spacetime were calculated in \cite{birrell,pavao}:
\begin{equation}
\label{solution}
u(x,k) = \frac{\left( i + \frac{R \omega}{a(t)} \right)}{(2 \pi \omega)^{\frac{3}{2}}\sqrt{2R^2}} e^{i \frac{R \omega}{a(t)}} e^{i \mathbf{r} \cdot \mathbf{k}},
\end{equation}
with $|\mathbf{k}| = \omega$.
Here, $R$ is the radius of the de Sitter Universe and it relates to the Hubble parameter ($H$) through
\begin{equation}
R = \frac{1}{H}.
\end{equation}
Making use of spherical symmetry we expand the spatial part of the modes in Spherical Harmonics, such that \eqref{solution} reduces to
\begin{equation}
\label{sphericalsol}
u_{l m}(x,k) = 4 \pi i^l \frac{\left( i + \frac{R \omega}{a(t)} \right)}{(2 \pi \omega)^{\frac{3}{2}}\sqrt{2R^2}} e^{i \frac{R \omega}{a(t)}} j_l(\omega r) {Y_l^m}^* (\alpha, \beta) Y_l^m (\theta,\phi),
\end{equation}
with
\begin{equation}
u(x,k) = \sum_{l=0}^{\infty} \sum_{m=-l}^{l} u_{l m}(x,k).
\end{equation}
Here $\mathbf{k} = (\omega, \alpha, \beta)$ and $\mathbf{r} = (r,\theta,\phi)$.
Since \eqref{sphericalsol} also represent valid field modes we can test the entanglement between specific $l$ and $m$, thus greatly simplifying the problem. We therefore define a new Hadamard function $H_{lm;l'm'} (x,x')$ such that,
\begin{equation}
H(x,x') = \sum_{l=0}^{\infty} \sum_{m=-l}^{l} \sum_{l'=0}^{\infty} \sum_{m'=-l'}^{l'}  H_{lm;l'm'} (x,x').
\end{equation}
To simplify matters we restrict the calculation of entanglement to the case of two modes with $l=l'=0$ and with $a(t=t_{\textup{now}})=1$. In this way we get the following Hadamard function:
\begin{equation*}
H_{00;00} (r,r') = 4 \sqrt{2 \pi} \int d\omega \int d\omega' \Bigg (\frac{\sin(\omega r) \sin(\omega' r')}{2 R^2 \sqrt{\omega \omega'} r r'} \times
\end{equation*}
\begin{equation}
\label{hadamard0-1}
\times \Bigg\{ \left(R \omega +i\right) \left(R \omega'-i \right)e^{i R \left( \omega-\omega'\right)}+\left(R \omega -i\right) \left(R \omega'+i \right)e^{-i R \left( \omega-\omega'\right)} \Bigg\} \Bigg ),
\end{equation}
Defining the following functions,
\begin{subequations}
\label{Ifunctions}
\begin{align}
& I_1 (r) = \int_{\omega_{\textup{min}}}^{\omega_{\textup{max}}}dz \sqrt{z} \sin (r z) \sin(R z), \\
& I_2 (r) = \int_{\omega_{\textup{min}}}^{\omega_{\textup{max}}}dz \sqrt{z} \sin (r z) \cos(R z),\\
& I_3 (r) = \int_{\omega_{\textup{min}}}^{\omega_{\textup{max}}}dz \frac{\sin (r z)}{\sqrt{z}} \sin(R z),\\
& I_4 (r) = \int_{\omega_{\textup{min}}}^{\omega_{\textup{max}}}dz \frac{\sin (r z)}{\sqrt{z}} \cos(R z), \\
& I_5 (r) = \int_{\omega_{\textup{min}}}^{\omega_{\textup{max}}}dz \sqrt{z^3} \sin (r z) \sin(R z), \\
& I_6 (r) = \int_{\omega_{\textup{min}}}^{\omega_{\textup{max}}}dz \sqrt{z^3} \sin (r z) \cos(R z),
\end{align}
\end{subequations} 
equation \eqref{hadamard0-1} becomes
\begin{equation}
\label{hadamard0}
\begin{split}
H_{00;00} (r,r') = \frac{4 \sqrt{2 \pi}}{r r'} \Bigg \{ I_1 (r) \left(I_1(r') + \frac{I_4 (r')}{R} \right)+I_2 (r) \left(I_2 (r') - \frac{I_3 (r')}{R} \right)+\\
+\frac{I_3 (r)}{R} \left(\frac{I_3(r')}{R}-I_2 (r') \right)+\frac{I_4 (r)}{R} \left( \frac{I_4 (r')}{R}+I_1 (r')\right)\Bigg \}.
\end{split}
\end{equation}
In \eqref{Ifunctions}, $\omega_{\textup{max}}$ and $\omega_{\textup{min}}$ are the UV and IR cutoffs, respectively. Here, we take them to be
\begin{equation}
\left\{\begin{matrix}
\omega_{\textup{max}} = l_p^{-1},  \ \\ 
\omega_{\textup{min}} = R^{-1},
\end{matrix}\right.
\end{equation}
where $l_p$ is the Planck length.\\
In a similar way we get
\begin{equation}
\begin{split}
\partial_t H_{00;00} (r,r') = -\frac{4 \sqrt{2 \pi}}{r r'} \Bigg \{\frac{I_6 (r) I_4 (r')+ I_5 (r) I_3 (r')}{R} - I_5(r) I_2 (r') + I_6(r) I_1(r')\Bigg \},
\end{split}
\end{equation}
\begin{equation}
\begin{split}
\partial_{t'} H_{00;00} (r,r') = -\frac{4 \sqrt{2 \pi}}{r r'} \Bigg \{\frac{I_6 (r') I_4 (r)+ I_5 (r') I_3 (r)}{R} - I_5(r') I_2 (r) + I_6(r') I_1(r)\Bigg \},
\end{split}
\end{equation}
and
\begin{equation}
\begin{split}
\partial_t \partial_{t'} H_{00;00} (r,r') = \frac{4 \sqrt{2 \pi}}{r r'} \Bigg \{I_6 (r) I_6 (r') + I_5 (r) I_5(r')\Bigg \}.
\end{split}
\end{equation}
Now, the functions $I_j(r)$ are solvable in the limits of $r \rightarrow R$ and $r \rightarrow 0$, very near and very far from the de Sitter Horizon. These solutions are shown in appendix \ref{App:A}.

With the Hadamard functions and their derivatives we are ready to test for the existence of entanglement near and far from the Horizon. Now we only need to choose the type of box in which the fields are averaged.

As stated before, cubic boxes can be used. Since in the present work we are also interested in the effects that the existence of a de Sitter Horizon has on entanglement, we can simulate such a horizon by averaging $\hat{\phi}$ inside a spherical shell of outer radius $R$ and thickness $l_p$.

Let us consider, in a more general way, a spherical shell of outer radius of $R_{\textup{out}}= r+\frac{l}{2}$ and a thickness of $l$, such that $l \ll R_{\textup{out}}$ and $V \simeq 4 \pi r^2 l$. We test the validity of these boxes by using the same method as in \cite{kimbound}:
\begin{equation}
\hat{\Phi}_l(r) = \frac{1}{r^2 l} \int_{r-\frac{l}{2}}^{r+\frac{l}{2}} du \ u^2 \hat{\phi}(u).
\end{equation}
Expanding $\hat{\phi}(u)$ around $u=r$ we get
\begin{equation}
\hat{\phi}(u) = \sum_j \frac{(u-r)^j}{j!} \partial_u^j \hat{\phi}(r),
\end{equation}
which means that
\begin{equation*}
\hat{\Phi}_l(r) = \sum_j \frac{1}{r^2 l} \frac{\partial_u^j \hat{\phi}(r)}{j!} \int_{r-\frac{l}{2}}^{r+\frac{l}{2}} d u \  u^2(u-r)^j =
\end{equation*}
\begin{equation*}
=\sum_j \frac{1}{r^2 l} \frac{\partial_u^j \hat{\phi}(r)}{(j+3)!} r^2 (j^2+5j+6)\times \begin{cases}
l^{j+1} & \text{ if } j \textup{ is even} \\ 
0 & \text{ if } j \textup{ is odd}
\end{cases}=
\end{equation*}
\begin{equation}
= \sum_j  \frac{\partial_u^j \hat{\phi}(r)}{(j+3)!} (j^2+5j+6) l^{2j} \Leftrightarrow \lim_{l \rightarrow 0}\hat{\Phi}_l(r) = \hat{\phi}(r),
\end{equation}
where we used $r-l \simeq r$ in the second line.
 
This means that we can test for the entanglement between two cubic boxes, one cubic box with one spherical shell (which, if $R_{\textup{out}} = R$, represents the de Sitter Horizon) or between two spherical shells.

\section{Numerical Results}

\begin{figure}[!bht]
\centering
\includegraphics[scale = 0.6]{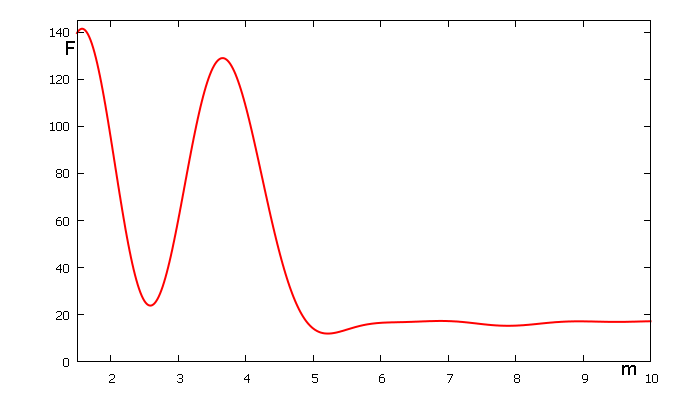}
\caption{Plot of $F$ vs $m$ (position of second state $B'$), with $n=0.5$, so the first state $B$ is at the Horizon. The position of the second state varies from $m= 1.5$ to $m=10$. For $m>5$, $F$ tends asymptotically to a positive constant.}
\label{fig:SS1}
\end{figure}

\begin{figure}[!bht]
\centering
\includegraphics[scale = 0.6]{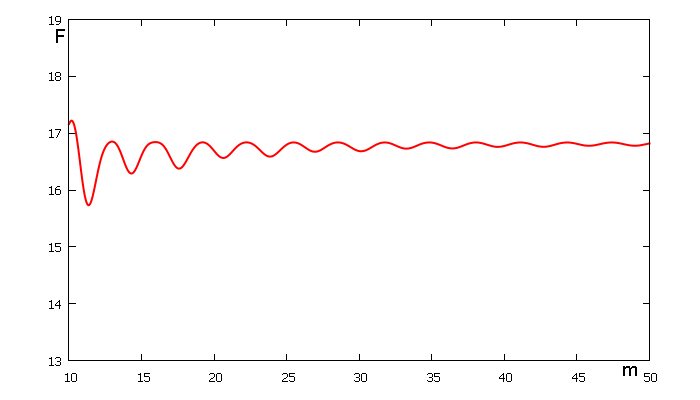}
\caption{Plot of $F$ vs $m$, with $n=0.5$ to represent the Horizon. The second state is varied from $m=10$ to $m=50$. The amplitude of the oscillations of $F$ decreases with increasing $m$ and $F$ tends to a constant.}
\label{fig:SS4}
\end{figure}

The behavior of the entanglement, via $F$, eq. \eqref{simoncrit}, between two spherical shells ($B$ and $B'$) near the Horizon is shown in figures \ref{fig:SS1} and \ref{fig:SS4}. Since the boxes are very near the Horizon, we parametrize the position of each box in terms of the distance to the mathematical Horizon\footnote{We define mathematical Horizon as the position $r=R$, and the physical Horizon (or just Horizon) is the spherical shell of Planck length thickness proposed in \cite{weinberg1,pavao}.}: the position $r$ of $B$ in terms of the variable $n$ and the position $r'$ of $B'$ in terms of $m$, as
\begin{subequations}
\begin{align}
&r = R-n l_p,\\
&r' = R-m l_p.
\end{align}
\end{subequations}
One of the shells (say $B$) is at a fixed distance with $n=0.5$, such that it represents the states of the Horizon. The position of other shell ($B'$) varies with $m$ in order to study the presence of entanglement between vacuum states $B'$ near the Horizon and vacuum states $B$ in the Horizon itself. The results were obtained using Planck units (where $l_p=1$).

The figures show that $F(n=0.5,m)>0$ (for all space where $R-r \ll R$) which means that entanglement between states near the Horizon and in the Horizon is found. The function $F(0.5,m)$ oscillates and eventually tends to a constant, $F (0.5,0.5 \ll m \ll \frac{R}{l_p})\simeq 16.84$ (where it remains for very large $m$). Entanglement is also detected between the Horizon and a cubic box of volume $V'=l_p^3$. In that case, $F(0.5,m) \simeq 16.84$ always. This behavior results from the fact that, if both boxes are spherical shells ($V=V'$) the correlations associated to the box $B'$ with itself ($\text{Det}(\mathbf{B})$) and the correlations between $B$ and $B'$ ($\text{Det}(\mathbf{C})$ and $\text{Det}(\mathbf{V})$) tend to zero when $m$ gets bigger such that \eqref{simoncrit} becomes
\begin{equation}
F(0.5,m \rightarrow \infty) \rightarrow \textup{Det}(\mathbf{A})(0.5,m)-\frac{1}{4},
\end{equation}
the correlations of box $B$ dominate $F$ (as can be seen in figures \ref{fig:SS1} to \ref{fig:SS4}).
However, if the second box is a cubic box ($V \gg V'$) , then $H(r',r')$ and $H(r,r')$ are always smaller than $H(r,r)$, such that
\begin{equation}
F(0.5,m) = \textup{Det}(\mathbf{A})(0.5,m)-\frac{1}{4},
\end{equation}
for any $m$ in the valid region.
In both cases, entanglement is always detected. However, in both cases, the states are not Gaussian states, since \eqref{uncrt} implies that $\textup{Det}(\mathbf{V})>0$ \cite{rsimon}, which is not fulfilled for any values of $n$ or $m$ tested. 

\begin{figure}[!bht]
\centering
\includegraphics[scale = 0.6]{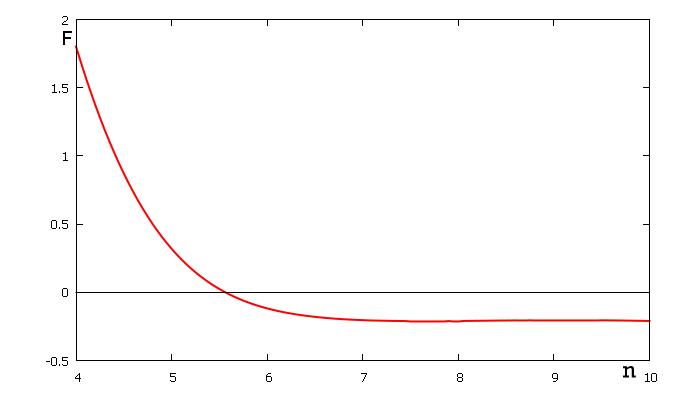}
\caption{Plot of $F$ vs $n$ (position of the first state $B$), with second state $B'$ at $m=10 000$. Entanglement ceases to be detected around $n\simeq 5.5$ which means that the rest of the Universe is only entangled with states very near the mathematical Horizon ($R$).}
\label{fig:SS5}
\end{figure}

When the box $B'$ is very far from the  horizon, $m \gg 0.5$ (in this case we set $m=10 000$), we can move the position of box $B$ to see when entanglement ceases. The results are shown in figure \ref{fig:SS5}. As we can see, $F(n,m>6)<0$ from $n \sim 5.5$ on, which means that from that point on we cannot determine if entanglement between $B$ and $B'$ exists. Entanglement between the rest of the Universe (regions where $F \simeq \textup{const}$) and a shell near the Horizon is only detected for $n \sim O(1)$. This seems to suggest that, if we consider the states belonging to the Event Horizon of de Sitter space as all the states between the $R$ and $R-l_p$, then most of the entanglement detected is between states of the Event Horizon and the rest of the Universe. Or, if we assume a priori that entanglement should exist between the states in the Horizon and the rest of the Universe (like in \cite{weinberg1}), then these results could be taken as an indication that the states belonging to the Horizon must fit in a spherical shell of thickness $\sim 5 l_p$. 

We can also analyze entanglement of states very far the Horizon ($ r \ll R$). In this limit the functions in \eqref{Ifunctions} rapidly tend to zero but in \eqref{hadamard0} we have a factor of $r^{-1}$ multiplying each function $I_j$ such that a divergence may arise. To evaluate the behavior of $H_{00;00}(r,r')$ we expand the functions $I_j$ up to second order,
\begin{equation}
I_j(r) r^{-1} \simeq \alpha_j r^{-1} + \beta_j + \gamma_j r.
\end{equation}
The results of this expansion are shown in appendix \ref{App:B}. As can be seen there, $\alpha_j = \gamma_j = 0$ which means that near Earth ($r \rightarrow 0$), the Hadamard functions are constant - the correlations between two states very far from the Horizon do not depend on their positions (unless the volumes of the boxes chosen to average the states depend on the position of the states).

The terms $\beta_j$ are very small, which means that
\begin{equation}
F(n,m) \simeq -\frac{1}{4}
\end{equation} 
for $n,m \gg 5$, independent of the shapes of the boxes (spherical shells or cubic). Also,
\begin{equation}
F(n,m) \simeq \textup{Det}(\mathbf{A})-\frac{1}{4}
\end{equation} 
for $n \lesssim 5$ and $m \gg 5$, as long as $B$ is a spherical shell of thickness $l_p$.

Thus, no entanglement is detected between states far away from the Horizon. Because of the weakness of the correlations of states very far from the Horizon, $F$ acquires the same positive value when one spherical shell is close to the horizon, $r \gtrsim R-5l_p$, for any choice of the other state as long as $r' \ll R$ .

\section{Conclusion} 
Studying the entanglement of states near and far from the Horizon of a de Sitter Universe we find that all states in the Universe are entangled with states very close to the horizon (less than 10 Planck lengths, a value consistent with ref. \cite{weinberg1}). As quantum correlations reduce the entropy of the system \cite{wolf}, this observation lends support to the holographic principle and provides a mechanism to resolve the issue of overcounting vacuum states which is at the root of 
the Cosmological Constant problem. All information in the Universe is connected with information at the Horizon, as stated in the Holographic Principle and in agreement with previous work  \cite{weinberg1,marks,shinji,bombelli,muller,wolf}.

It is worth noticing that the results presented in our study are obtained in a de Sitter Universe. Our Universe is described by a Friedmann metric that will eventually evolve into a de Sitter metric, but in the present epoch the results will differ somewhat. However, since our Universe is already dominated by Dark Energy (which constitutes about 70\% of the total density of the Universe), a de Sitter metric is a good approximation to our current metric and the results we obtain are expected to be  reasonable.

Another important property of our analysis is that our states are not gaussian, which means that $F$ is not a quantitative measure of entanglement. It only proves that entanglement exists when $F$ is positive and it does not disprove entanglement when $F$ is negative. However our expectation is that $F$ can be considered an approximate quantitative measure of entanglement in the situation of non-gaussian states.

Given the lack of a method to determine uniquely the separability criteria of non-gaussian continuous variable states, work is in progress to obtain a clearer quantitative measure of entanglement in order to shed more light on its role in the Holographic Principle and in the resolution of the Cosmological Constant problem.

\appendix
\section{The Integrals $I_i(r)$ close to the Horizon}
\label{App:A}

The functions $I_j(r)$ take the following form, for $r \rightarrow R$:
\begin{subequations}
\begin{align}
\label{corr1}
& I_1(r) \simeq -\sqrt{\frac{\pi}{8}} \frac{1}{(R-r)^{\frac{3}{2}}} \left\{S\left(\sqrt{\frac{2(R-r)}{\pi l_p}}\right)- \sqrt{\frac{2(R-r)}{\pi l_p}} \sin \left(\frac{R-r}{l_p}\right) \right\},\\
\label{corr2}
& I_2(r) \simeq -\sqrt{\frac{\pi}{8}} \frac{1}{(R-r)^{\frac{3}{2}}} \left\{C\left(\sqrt{\frac{2(R-r)}{\pi l_p}}\right)-\sqrt{\frac{2 (R-r)}{\pi l_p}} \cos\left(\frac{R-r}{l_p}\right)  \right\},\\
\label{corr3}
& I_3(r) \simeq \sqrt{\frac{\pi}{2 (R-r)}} C\left(\sqrt{\frac{2(R-r)}{\pi l_p}}\right),\\
\label{corr4}
& I_4(r) \simeq -\sqrt{\frac{\pi}{2 (R-r)}} S\left(\sqrt{\frac{2(R-r)}{\pi l_p}}\right),\\
\label{corr5}
& I_5(r) \simeq \frac{3}{2}\sqrt{\frac{\pi}{8}} \frac{1}{(R-r)^{\frac{5}{2}}} \Bigg\{ -C\left(\sqrt{\frac{2(R-r)}{\pi l_p}}\right)+\sqrt{\frac{2(R-r)}{\pi l_p}} \cos\left(\frac{R-r}{l_p}\right)+\frac{1}{3}\sqrt{\frac{8}{\pi}} \left(\frac{R-r}{l_p}\right)^{\frac{3}{2}} \sin\left(\frac{R-r}{l_p}\right)\Bigg \},\\
\label{corr6}
&I_6(r) \simeq \frac{3}{2}\sqrt{\frac{\pi}{8}} \frac{1}{(R-r)^{\frac{5}{2}}}  \Bigg\{ S\left(\sqrt{\frac{2(R-r)}{\pi l_p}}\right)-\sqrt{\frac{2 (R-r)}{\pi l_p}} \sin\left(\frac{R-r}{l_p}\right)+\frac{1}{3}\sqrt{\frac{8}{\pi}} \left(\frac{R-r}{l_p}\right)^{\frac{3}{2}} \cos\left(\frac{R-r}{l_p}\right) \Bigg \}.
\end{align}
\end{subequations}
$S(x)$ and $C(x)$ denote the Fresnel integrals,
\begin{subequations}
\begin{align}
& S(x) =\int_{0}^{x} \sin(t^2) dt\\
& C(x) = \int_{0}^{x} \cos(t^2) dt
\end{align}
\end{subequations}
\linebreak
\section{The Integrals $I_i(r)$ far from the Horizon}
\label{App:B}

The functions $I_j(r)$ take the following form, for $r \ll R$ and using the fact that $R\gg l_p$:
\begin{subequations}
\begin{align}
& I_1(r) \simeq -\frac{1}{\sqrt{l_p}}\frac{r}{2 R^{2}} \left\{3 \sin \left(\frac{R}{l_p}\right)+2\left(\frac{R}{l_p}\right) \cos
   \left(\frac{R}{l_p}\right)\right\}, \\
& I_2(r) \simeq \frac{1}{\sqrt{l_p}}\frac{r}{2 R^{2}} \left\{3 \cos \left(\frac{R}{l_p}\right)+2 \left(\frac{R}{l_p}\right) \sin
   \left(\frac{R}{l_p}\right)\right\}, \\
& I_3(r) \simeq -\frac{1}{\sqrt{l_p}}\frac{r}{R} \cos \left(\frac{R}{l_p}\right),\\
& I_4(r) \simeq \frac{1}{\sqrt{l_p}}\frac{r}{R} \sin \left(\frac{R}{l_p}\right),\\
& I_5(r) \simeq \frac{1}{\sqrt{l_p}}\frac{r}{4 R^{3}} \left\{10 \left(\frac{R}{l_p}\right) \sin \left(\frac{R}{l_p}\right)+ \left[15-4 \left(\frac{R}{l_p}\right)^2\right] \cos \left(\frac{R}{l_p}\right)\right\}, \\
& I_6(r) \simeq \frac{1}{\sqrt{l_p}}\frac{r}{4 R^{3}} \left\{10 \left(\frac{R}{l_p}\right) \cos \left(\frac{R}{l_p}\right)- \left[15-4 \left(\frac{R}{l_p}\right)^2\right] \sin \left(\frac{R}{l_p}\right)\right\}.
\end{align}
\end{subequations}

\acknowledgments

We would like to thank Jorge Morais for many useful discussions.

\bibliographystyle{unsrt}
\bibliography{references}

\end{document}